\documentclass[12pt]{article}
\usepackage[english]{babel}
\usepackage{times}
\usepackage[T1]{fontenc}
\usepackage{epsfig}

\usepackage{amsmath}
\usepackage{amssymb}

\begin{document}

\title{Note on $Spin(3,1)$ tensors, the  Dirac field  \\ and  $GL(k, \mathbb{R})$ symmetry}

\author{H. Arod\'z$\:^a$  \; and \; Z. \'Swierczy\'nski$\:^b$ \\ 
	{\small $^a$  Institute of Theoretical Physics,
Jagiellonian University, Cracow, Poland} \footnote{henryk.arodz@uj.edu.pl} \\ {\small $^b$ Pedagogical University, Cracow, Poland} \footnote{zs@up.krakow.pl} }
\date{$\;$}

\maketitle

\vspace*{0.3cm}

\begin{abstract}
We show that the rank decomposition of a real matrix $r$, which is a $Spin(3,1)$ tensor, leads to  $2k$   Majorana bispinors, where $k= rank\: r$. The Majorana bispinors are determined up to  local $GL(k, \mathbb{R})$ transformations. The bispinors are combined in pairs to form $k$ complex Dirac fields.  We analyze in detail the case  $k=1$, in which there is just one Dirac field with the well known standard Lagrangian.  The  $GL(1, \mathbb{R})$ symmetry  gives rise to a new conserved current, different from the well known $U(1)$ current. The $U(1)$ symmetry is present too. All global continuous internal symmetries in the $k=1$ case form the $SO(2,1)$ group.   As a side result, we  clarify the  discussed in literature  issue whether there exist algebraic  constraints for the  matrix $r$ which would be equivalent to the condition $rank\: r=1$.

\end{abstract}

\vspace*{1cm}
\thispagestyle{empty}

\pagebreak
\setcounter{page}{1}
\section{ Introduction}

Let us begin with some historical background for our work. Probably the first appearance of a complex four-component bispinor was as the wave function of the relativistic Dirac particle.  It was introduced by P. A. M. Dirac in a rather mathematical way, in a search for concrete matrix representations of the formal algebra of his  $\gamma^{\mu}$ matrices. Furthermore, it turns out that the Poincar\'e transformations of the Minkowski space-time \footnote{We consider fields on the four-dimensional Minkowski space-time $M$ with the metric $\eta = \mbox{diag}(1, -1,-1,-1)$. The Dirac matrices obey the relations $\gamma^{\mu}\gamma^{\nu} + \gamma^{\nu}\gamma^{\mu} = 2 \eta^{\mu\nu} I_4$,  where $I_4$ denotes the four by four unit matrix} are represented in the linear space of the bispinors by four by four matrices forming the  $\mbox{Spin}(3,1)$  group. In consequence, the bispinor wave function has a rather weird feature that it changes sign after spatial rotations of reference frame by $2\pi$. It is clear that even though there is a  firm mathematical basis  for the bispinor wave function, it would be desirable to provide  alternative motivations for it, or perhaps even a substitute. Certain efforts in this direction are based on the observation that expectations values of observables for the Dirac particle involve quadratic bilocal expressions of the form $\Psi^*_{\alpha}(x) \Psi_{\beta}(y)$, where $\Psi_{\alpha}$ are components of the Dirac bispinor,  $ x, y \in M$, and $^*$ denotes the complex conjugation. Such expressions are in fact  matrix elements of  density operator.  They are invariant under the rotations by $2\pi$. Moreover, they belong to a reducible representation of the Poincar\'e group, which can be decomposed into more familiar bosonic scalar and vector components. These facts motivated formulations of the quantum mechanics of the Dirac particle solely in terms of the local quadratic expressions  $\Psi^*_{\alpha}(x) \Psi_{\beta}(x)$.  In some sense, such  attempts were successful,  but the new formulations turned out rather cumbersome, and they did not replace the original Dirac's one, especially in applications.  For examples of works in this direction see, e.g., \cite{1},  \cite{2},  \cite{3}, \cite{4}, \cite{5}, and references therein.

The present work is inspired by those papers, but in actual fact there are crucial differences. 
 We start from expectation values of products of two  quantized Hermitian Majorana fields in arbitrary quantum state. Such expectation values in general do not have the product form mentioned above. Instead, they form a generic four by four real matrix  $r(x)$. We show that the rank decomposition of such a matrix leads to one or several independent Dirac bispinors, depending on the rank of the matrix.  Thus, we  derive the Dirac bispinor, not replace it. 
Moreover, we work in the framework of field theory. In our paper   $\Psi(x)$ is the classical Dirac field, and not the wave function of the particle. Generally speaking, we offer a new perspective on the classical Dirac field. 

Let us now describe the content of our paper in more detail.  We recover the classical Dirac field (or fields) as a kind of coordinates in nonlinear Lorentz invariant subspaces of the space of  matrix valued functions $r(x),$ $x\in M$. These functions are introduced as  the expectation values of products of two free quantized Majorana fields, see formula (4) below.   The matrix elements $r_{\alpha \beta},$  
 $\alpha, \beta = 1,2,3,4,$ are real. The invariant subspaces are defined by fixing the rank of matrices $r(x)$. Such unusual derivation of the  Dirac field brings an unexpected  bonus  in the form of an internal $GL(k, \mathbb{R})$ symmetry, with the corresponding conserved current. The symmetry is present irrespectively of the mass of the field. To the best of our knowledge, it is a new internal symmetry for the Dirac field. The standard well known $U(1)$ symmetry also is present.    The number of the Dirac fields is equal to the rank of the matrices $r(x)$.  In the most general case, i.e.,  $rank \: r(x)=4$, we 
 obtain a  multiplet of four independent Dirac fields. We consider in detail the case of $rank \: r(x)=1$.  In particular, we compare the 
 conserved $GL(1, \mathbb{R})$ and $U(1)$ currents, and we introduce the covariant derivative and the gauge field in the case of local $GL(1, \mathbb{R})$ group. We find three independent global internal symmetries which together form the $SO(2,1)$ group.

 The plan of our paper is as follows. 
  In the next section we introduce the space of real matrix functions $r(x)$, we discuss their relativistic transformations,  and we point out the nonlinear Lorentz invariant subspaces. Section 3 is devoted to the discussion of the rank one case, in which there is just one Dirac field.  Here we describe the new internal symmetry $GL(1, \mathbb{R})$,   the corresponding conserved current, and the covariant derivative pertinent to the local version of this symmetry.  
   The cases of higher rank   are briefly addressed in Section 4. Section 5 contains a summary and  remarks. In the Appendix A we recall certain mathematical facts about rank of matrices, for convenience of the reader.  
In the Appendix B we prove that the constraints (12) imply that the rank of matrix $r$ is equal to 1 or 0.

\section{ The tensor representations of the $Spin(3,1)$ group}

In this Section we introduce  the tensor representations of the $Spin(3,1)$ group considered in our paper.  
We prefer the standard description of bispinors and $\gamma^{\mu}$ matrices, as
found in  textbooks on theory of fields, e.g., \cite{6}, \cite{7}, as opposed to  much more formal mathematical framework known as the theory of Clifford algebras and their representations, see e.g. \cite{1}.

The tensors are introduced as expectation values of products of quantized free Majorana fields. 
The Majorana quantum field $\hat{\psi}(x)$ is Hermitian.  It has the following relativistic transformation law 
\begin{equation} U^{-1}(L) \hat{\psi}(x) U(L) = S(L) \hat{\psi}(L^{-1}x),   \end{equation} 
where $L$ denotes arbitrary proper ortochronous Lorentz transformation, and $U(L)$ is a unitary representation of the Lorentz transformation in the Fock space pertinent for the field. 
 The four by four matrix $S(L)$ has the form
\begin{equation} 
S(L) = \exp(b_{\mu\nu}[\gamma^{\mu}, \gamma^{\nu}]/8).  
\end{equation}
Here $ b_{\mu\nu}$ are real coefficients, $ b_{\mu\nu} = - b_{\nu\mu}$, and $[\; ,\; ]$ denotes the commutator of  matrices.  The Dirac matrices $\gamma^{\mu}$  are taken in one of the Majorana representations, i.e., they are purely imaginary. Moreover,   $\gamma^{0\:T} = - \gamma^{0}, \; \gamma^{i\:T}= \gamma^{i}$, where $T$ denotes the matrix transposition. It follows that also the matrix $\gamma_5 = i \gamma^0 \gamma^1 \gamma^2 \gamma^3$ is purely imaginary, and $\gamma_5^T = - \gamma_5$. It is clear from formula (2) that the matrices $S(L)$ are real.  They
form the group $Spin(3,1)$. The quantized free Majorana field is discussed in many works, see, e.g., \cite{8}, \cite{9}. We follow the notation from the paper \cite{9}.

The tensors we are going to investigate can be obtained, for example, as
 the expectation values  
\begin{equation}
s_{\alpha \beta}(x) = i \langle \Phi| \hat{\psi}_{\alpha}(x) \hat{\psi}_{\beta}(x) |\Phi \rangle,
\end{equation}
and
\begin{equation}
r_{\alpha \beta}(x) = i \langle \Phi| \hat{\psi}_{\alpha}(x) \hat{\chi}_{\beta}(x) |\Phi \rangle,
\end{equation}
where $\hat{\chi}(x)$ is another free quantized Majorana field  (anticommuting with $\hat{\psi}$), and
 $|\Phi\rangle$ is a certain state from the Fock space. We omit the standard discussion of how the local products of the quantized fields are defined because this point is irrelevant for our considerations. We shall exploit only algebraic properties of the tensors, passing over their possible physical applications. 
  Both tensors 
are real  \footnote{This is the reason for  including the coefficient $i$}.    $\:s_{\alpha \beta}(x)$ is  antisymmetric, $s_{\alpha \beta}(x) = - s_{\beta \alpha}(x)$, because the components of the Majorana field anticommute.
The tensor $r_{\alpha \beta}(x)$  does not have any definite symmetry properties. 
Both tensors have the dimension $\mbox{cm}^{-3}$ in the natural units.

The  transformation law of the tensors  follows from formulas (1), (3), and (4), 
\begin{equation}
s'_{\alpha\beta}(x) = S(L)_{\alpha\delta}  S(L)_{\beta\eta} \: s_{\delta\eta}(L^{-1}x),  \;\;\; r'_{\alpha\beta}(x) = S(L)_{\alpha\delta}  S(L)_{\beta\eta} \: r_{\delta\eta}(L^{-1}x).
\end{equation}
Here $s'(x)$ and $r'(x)$ are defined by formulas (3), (4) with the state $|\Phi \rangle$ replaced by  $|\Phi ' \rangle = U(L) |\Phi \rangle $. 
Formulas (5) in the matrix form  read
\begin{equation} 
s'(x) = S(L)\: s(L^{-1}x)\: S(L)^T, \;\;\;\;\; r'(x) = S(L)\: r(L^{-1}x) \:S(L)^T.
\end{equation}
It is convenient to introduce equivalent tensors
\[ \tilde{s}(x) = i\: s(x)\: \gamma^0,   \;\;\;\; \tilde{r}(x) = i\: r(x)\: \gamma^0, \]
which also are real. Note that $\tilde{s}$ is not antisymmetric, but \begin{equation} 
\tilde{s}^T(x) = \gamma^0 \tilde{s}(x) \gamma^0. 
\end{equation}
 The transformations of the new tensors have the form 
\begin{equation} 
\tilde{s}^{\:'}(x) = S(L)\: \tilde{s}(L^{-1}x)\: S(L)^{-1}, \;\;\;\;\; \tilde{r}^{\:'}(x) = S(L)\: \tilde{r}(L^{-1}x) \:S(L)^{-1}.
\end{equation}
Here we have used the well known formula 
\[ S(L)^T \gamma^0 = \gamma^0 S(L)^{-1}. \]
Formulas (8)  give two  representations of the $Spin(3, 1)$ group. They differ only by the space of pertinent matrices.  As is well known, these representations are reducible. The irreducible components are obtained by 
decomposing the matrices $\tilde{s}$,   $\tilde{r}$ in the well known matrix basis consisting of  sixteen real  matrices $ I_4,\: i\gamma_5,$  $  \:i \gamma^{\mu},\: \gamma_5 \gamma^{\mu},$  and $ [\gamma^{\mu},  \gamma^{\nu}]$.    Thus, 
\begin{equation}
\tilde{s}(x) = A(x) I_4 + i A_5(x) \gamma_5  + V_{5\mu}(x) \gamma_5 \gamma^{\mu},   
\end{equation}
and
\begin{equation}
 \tilde{r}(x) =  A(x) I_4 + i A_5(x) \gamma_5  + i V_{\mu}(x) \gamma^{\mu}  + V_{5\mu}(x) \gamma_5 \gamma^{\mu}  + S_{\mu\nu}(x) [\gamma^{\mu}, \gamma^{\nu}]. 
\end{equation}
Formula (9) takes into account the property (7). All fields present in these formulas are real-valued, and there are no  constraints for them.   With respect to Lorentz transformations, implemented by formulas (8),  the fields $A, A_5$ are scalars, $V_{\mu}$ and $V_{5\mu} $ 
four-vectors, and $S_{\mu\nu}$ is antisymmetric tensor.  When looking at  spatial rotations only, the $r(x)$ field contains four spin 0  and four spin one fields. Thus  the fields are bosonic. The sector with $S_{\mu\nu} \neq 0$, and all other fields vanishing, coincides with the adjoint representation of the $Spin(1,3)$ group. 

Note that the field $s(x)$ can be obtained from  $r(x)$ by imposing the condition of antisymmetry. For this reason, below we  shall discuss  only this latter field.

In literature,  a  lot of attention is given to a particular case of the tensor $r(x)$, namely
\[
r_{\alpha \beta}(x) = \psi_{\alpha}(x) \psi_{\beta}(x),
\]
where $\psi(x)$ is a real classical Majorana field, see, e.g., \cite{1},  \cite{2}, \cite{3}, \cite{4}, \cite{5} \footnote{In fact, these papers are focused mainly on the Dirac field, while we are interested in the Majorana field}. In the matrix form 
\begin{equation} r(x) = \psi(x) \psi(x)^T. \end{equation}
Such matrix $r$ is symmetric  and  singular, $rank\:r =1$. As mentioned in the Introduction, the main goal of those works is to replace the classical Majorana field (or rather the Dirac field) with the tensor field of the form (11) with certain modifications appropriate for the Dirac field. The  motivation is that the tensor field is closer to observables than the fermionic field itself. 
The matrices of rank one considered in the next Section are more general than 
 (11).

 The space of real 4 by 4 matrices $\tilde{r}(x)$,  or $r(x)$, is equivalent to $R^{16}$ (at arbitrary fixed point $x\in M$). As said above, its linear subspaces  defined in the decomposition (10) are invariant with respect to transformations (8). 
 Generalizing  the considerations of matrices of the form (11)  found in literature, below we study another set of invariant subspaces. They are formed  by all matrices  $\tilde{r}(x)$ of the same rank. For convenience of the reader, we recall basic facts about rank of matrices in the Appendix  A.  The matrix $r(x)$ has the same rank as  $\tilde{r}(x)$, hence they both belong to the same subspace. The important point is that  such subspaces are invariant with respect to the Poincar\'e transformations because  the rank is invariant with respect to  transformations (8) or (6).   Note that these subspaces are not linear because linear combination of matrices of the same rank can have a different rank.

\section{The matrices $r(x)$ of maximal rank one}

In the present Section we consider only matrices $r(x)$ of the rank equal to 1 or 0. We prove in the Appendix B that this condition is equivalent to the following  Lorentz invariant constraints for $\tilde{r}$
\begin{equation}
\tilde{r}(x) \gamma^0 \tilde{r}(x)^T = 0, \;\;\; \tilde{r}(x) \gamma_5\gamma^0 \tilde{r}(x)^T = 0, \;\;\; \tilde{r}(x) \gamma^{\mu} \gamma_5\gamma^0 \tilde{r}(x)^T = 0. 
\end{equation}
They imply a number of quadratic constraints for the bosonic fields introduced in formula (10). 
For instance, the first of conditions  (12) gives the following six relations \footnote{\;$\epsilon^{\alpha\beta\mu \nu}$ is the totally antisymmetric symbol, $\epsilon^{0123}=+1$}  
\[ A^2 - A_5^2   +V^{\mu} V_{\mu}  - V_5^{\mu} V_{5\mu} + 8 S^{\mu\nu}S_{\mu\nu} =0, \;\;
A A_5 - V^{\mu} V_{5\mu} + 2 \epsilon^{\alpha\beta\mu \nu} S_{\alpha\beta}S_{\mu\nu}=0, \]
\[ A V_{5\mu} + A_5 V_{\mu} + 4 V_5^{\nu} S_{\mu\nu} + 2 \epsilon_{\mu}^{\;\;\nu\alpha\beta}V_{\nu} S_{\alpha\beta} =0. \]

 Directly from the definition of the rank, see the Appendix  A,  it follows
that the matrix $r(x)$ of the rank 1 or 0 can be written in the form
\begin{equation}
r(x) = \left( \begin{tabular}{c} $ \psi_1(x)$ \\ $\psi_2(x)$ \\ $\psi_3(x)$ \\ $\psi_4(x)$ \end{tabular} \right) \left(\chi_1(x), \chi_2(x), \chi_3(x), \chi_4(x)\right),
\end{equation}
or  concisely  $r(x) = \psi(x) \chi(x)^T$, where $\psi(x)$ and $\chi(x)$ are independent four-component Majorana bispinors. At a given point $x$, the rank of such matrix $r(x)$ is equal to 0  when $\psi(x)=0$ or $\chi(x)=0$, i.e., when $r(x)=0$.  Both Majorana bispinors in formula (13) are real.  Their Lorentz transformations are assumed to  have the form
\[ \psi'(x) = S(L)\: \psi(L^{-1}x), \;\;\;  \chi'(x) = S(L)\: \chi(L^{-1}x), 
\]
in accordance with transformation (6) of matrix $r(x)$. 
From these Majorana bi\-spinors we can construct the complex Dirac field $\Psi(x)$, 
\begin{equation}
\Psi(x) = \psi(x) + i\: \chi(x).
\end{equation}

For given field $r(x)$, formula  (13) leaves the freedom of local rescaling the two Majorana bispinors. Namely, the fields 
\begin{equation}
\psi_a(x) = a(x) \psi(x), \;\;\; \chi_a(x) = a(x)^{-1} \chi(x), 
\end{equation}
where $a(x)$ is a dimensionless real scalar function, and $a(x) \neq 0$, give the same $r(x)$ as $\psi$ and $\chi$. At each fixed $x\in M$  transformations (15) form  one dimensional general linear group, denoted as $GL(1,\mathbb{R})$. The Dirac field is transformed as follows
\[
\Psi_a(x) =  a(x) \psi(x) + i \: a(x)^{-1} \chi(x). 
\]
Writing $a(x)= \exp(\alpha(x))$, and $\Psi_{\alpha}$ instead of $\Psi_a$, we obtain
\begin{equation}
\Psi_{\alpha}= \cosh\alpha\: \Psi + \sinh\alpha\: \Psi^*,   \;\;\; \Psi_{\alpha}^*= \sinh\alpha\: \Psi   +  \cosh\alpha\: \Psi^*,  
\end{equation}
where we have omitted the argument $x$. 
Note that the exponential parameterization assumes that $a(x) >0$. Negative $a(x)$ are obtained by flipping the sign of $\Psi$. This discrete transformation accompanies continuous transformations (16).

All Dirac fields $\Psi_{\alpha}(x)$ are equivalent in the sense that they give the same field $r(x)$. Note that we have found the local hyperbolic transformations (16), while the well known, and perhaps expected by the reader, $U(1)$ group is not present at this stage.

Thus far we have discussed only transformations of the fields. For a complete theory we need  equations of motion. They can be obtained from a Lagrangian built from relativistic invariants. As a  simple real scalar relativistic invariant we may take 
\[ {\cal L}_m = i m \:\mbox{Tr}\left(r(x)\gamma^0\right), \]
where $m$ is a constant (the mass coefficient), and $r(x)$ has the form (13).  
Simple calculations give
\begin{equation}
{\cal L}_m = i m \chi^T \gamma^0 \psi = - \frac{1}{2} m \: \Psi^{\dagger} \gamma^0 \Psi
= - \frac{1}{2} m \overline{\Psi} \Psi, 
\end{equation}
where $\overline{\Psi} = \Psi^{\dagger} \gamma^0$,  $\:^{\dagger}$ denotes the Hermitian conjugation, and $\mbox{Tr}$ denotes the trace of a matrix.  Formula (17) does not contain terms like $\Psi^T \gamma^0 \Psi$, etc., because they vanish due to antisymmetry of the matrix $\gamma^0$.   

Another relevant invariant, which yields the kinetic part of the Lagrangian, has the form
\[
{\cal L}_k = \mbox{Tr} \left((\gamma^{\mu} \partial_{\mu}\psi) \chi^T  \gamma^0    \right). 
\]
Introducing the Dirac field, formula (14), we have
\begin{equation}
{\cal L}_k = \frac{i}{4} \left(\overline{\Psi} \gamma^{\mu} \partial_{\mu} \Psi - \partial_{\mu} \overline{\Psi} \gamma^{\mu} \Psi   \right)  +  \frac{1}{2} 
\mbox{ Im} \left(  \Psi^T \gamma^0  \gamma^{\mu} \partial_{\mu} \Psi  \right). 
\end{equation}
The last term on the r.h.s. can be written as the four-dimensional divergence $ \partial_{\mu} \mbox{Im}(\Psi^T \gamma^0 \gamma^{\mu} \Psi)/4$,  because the matrices $\gamma^0\gamma^{\mu}$ are symmetric. Thus, it can be abandoned.
These two invariants together (and multiplied by 2) give the Dirac Lagrangian 
\begin{equation} 
{\cal L}_D = \frac{i}{2} \left(\overline{\Psi} \gamma^{\mu} \partial_{\mu} \Psi - \partial_{\mu} \overline{\Psi} \gamma^{\mu} \Psi   \right) - m \overline{\Psi} \Psi,
\end{equation}
known from textbooks.

Let us check internal symmetries of the Lagrangian ${\cal L}_D$. Its component ${\cal L}_m$ evidently is  invariant with respect to the local $GL(1,\mathbb{R})$
 transformations (15) and the related transformations (16). On the other hand, the term  ${\cal L}_k$ is not invariant under the local $GL(1,\mathbb{R})$ transformations because of the presence of derivatives.  Nevertheless,  it remains invariant with respect to the global ones. Also the kinetic part of the Dirac Lagrangian 
is invariant with respect to the global transformations of the form (16). 
The corresponding conserved current density is readily obtained from the Noether formula, 
\begin{equation}
j_s^{\mu}(x) = \frac{i}{2} \left( \Psi^{\dagger}\gamma^0 \gamma^{\mu} \Psi^* - \Psi^T \gamma^0 \gamma^{\mu} \Psi \right)= \mbox{Im}\left(\Psi^T\gamma^0\gamma^{\mu} \Psi \right) = 2\chi^T\gamma^0 \gamma^{\mu} \psi.
\end{equation}
The conserved total charge has the form  
\begin{equation}
Q_s = \frac{i}{2} \int \! d^3x \: \left( \Psi^{\dagger} \Psi^* - \Psi^T  \Psi \right) = \int \! d^3x \: \mbox{Im}\left( \Psi^T  \Psi \right).
\end{equation}

The Dirac Lagrangian (19) of course possesses also the well known global $U(1)$ invariance with transformations of the form
\begin{equation} 
  \Psi'(x) = \exp(i \beta) \:\Psi(x),
  \end{equation}
  where $\beta$ is real. 
 Interestingly, this symmetry has emerged on the level of Lagrangian -- there is no hint of it in formula (13), in contradistinction from the $GL(1,\mathbb{R})$ invariance.  Transformation (22) is equivalent to 
  \[ \psi'(x) = \cos\beta \:\psi(x) - \sin\beta \: \chi(x), \;\;\; \chi'(x) = \sin\beta \:\psi(x) + \cos\beta \:\chi(x), 
  \]
 and
 \[ r'(x) = \cos^2\beta\: r(x) - \sin^2\beta\: r(x)^T + \sin\beta \: \cos\beta\:\left(  \psi(x) \psi(x)^T - \chi(x) \chi(x)^T \right).
 \]  
Nevertheless, 
\[ \mbox{Tr}(r'(x) \gamma^0) =  \mbox{Tr}(r(x) \gamma^0),
\]
because the matrix $\gamma^0$ is antisymmetric, and therefore \[ \mbox{Tr}(r^T(x) \gamma^0) = - \mbox{Tr}(r(x) \gamma^0),  \;\; \mbox{Tr}(\psi \psi^T \gamma^0) =0, \;\; \mbox{Tr}(\chi \chi^T \gamma^0) =0. \]
The kinetic Lagrangian ${\cal L}_k$  changes under the $U(1)$ transformations (22) by a term which is a four-dimensional divergence. This can be seen also from formula (18): the last term on the r.h.s., which is the four-dimensional divergence, is not $U(1)$-invariant.  
Thus, in the presented derivation of the Dirac Lagrangian (19), the $U(1)$ invariance appears as a rather accidental symmetry, as opposed to the $GL(1,\mathbb{R})$ symmetry implied already by the basic formula (13).

The well known formulas for the  $U(1)$  current density and  conserved total charge for the Dirac field read
\[ j^{\mu}(x) = \overline{\Psi}(x) \gamma^{\mu} \Psi(x),  \;\; \; Q= \int\! d^3x\: \Psi^{\dagger}(x) \Psi(x). \]
Note that  the $U(1)$  charge $Q$ is not invariant under the  $GL(1,\mathbb{R})$
transformations (16). On the other hand, the charge $Q_s $, formula (21),  is not invariant with respect to the global  $U(1)$ transformations.
One can expect that in  quantum theory of the free Dirac field the corresponding operators will not commute with each other, and that also the commutator will be a  conserved charge. This topic of symmetries of the quantized Dirac field lies outside the scope of the present paper. 

The local $GL(1,\mathbb{R})$ invariance can be restored in the standard manner  by introducing appropriate gauge field  $B_{\mu}$ and  covariant derivatives, 
\begin{equation}
D_{\mu}(B) \psi = \partial_{\mu} \psi + B_{\mu} \psi, \;\;\; D_{\mu}(B) \chi = \partial_{\mu} \chi - B_{\mu} \chi.
\end{equation}
The gauge transformations have the form  (15) and
\begin{equation}
B^a_{\mu}(x) = B_{\mu}(x) - a^{-1}(x) \partial_{\mu} a(x). 
\end{equation}
The corresponding covariant derivative for  the Dirac field (14) is conveniently written in terms of eight-dimensional column formed by $\Psi(x)$ and $\Psi^*(x)$. The gauge transformation (16) gives
\begin{equation}
\left( \begin{array}{c} \Psi_{\alpha}(x) \\ \Psi^*_{\alpha}(x) \end{array} \right) = \left( \begin{array}{c c} I_4\: \cosh\alpha 
 & I_4\:\sinh\alpha  \\ I_4\:\sinh\alpha  & I_4\:\cosh\alpha \end{array} \right) \left( \begin{array}{c} \Psi(x) \\ \Psi^*(x) \end{array} \right),
\end{equation}
where on the r.h.s. there is 8 by 8 matrix formed from the four dimensional unit  matrix  $I_4$ multiplied by the hyperbolic functions.  
The covariant derivative reads
\begin{equation}
D_{\mu}(B) \left( \begin{array}{c} \Psi(x) \\ \Psi^*(x) \end{array} \right) = \partial_{\mu} \left( \begin{array}{c} \Psi(x) \\ \Psi^*(x) \end{array} \right) + B_{\mu}(x) \left(\begin{array}{cc} 0_4 & I_4 \\ I_4 & 0_4 \end{array}  \right)  \left( \begin{array}{c} \Psi(x) \\ \Psi^*(x) \end{array} \right).
\end{equation}
The term with $B_{\mu}$ field contains  8 by 8 matrix built from the four dimensional zero  $0_4$ and unit  $I_4$  matrices.

The Dirac Lagrangian (19) can easily be rewritten in terms of the eight-dimensional column formed by $\Psi(x)$ and $\Psi^*(x)$. Having the Lagrangian in this form, we 
replace the ordinary derivatives by  the covariant ones. It turns out that  the gauge field $B_{\mu}(x)$ is  coupled with the current $j_s^{\mu}(x)$ given by formula (20). 

For comparison, let us recall that the $U(1)$ gauge field $A_{\mu}(x)$ is introduced in the local $U(1)$ covariant derivatives, 
\begin{equation}
D_{\mu}(A) \left( \begin{array}{c} \Psi(x) \\ \Psi^*(x) \end{array} \right) = \partial_{\mu} \left( \begin{array}{c} \Psi(x) \\ \Psi^*(x) \end{array} \right) + i A_{\mu}(x) \left(\begin{array}{cc} I_4 & 0_4 \\ 0_4 & -I_4 \end{array}  \right)  \left( \begin{array}{c} \Psi(x) \\ \Psi^*(x) \end{array} \right). 
\end{equation}
In the Dirac Lagrangian, $A_{\mu}$ couples with the current $j^{\mu}(x)$. 

The $GL(1, \mbox{R})$ and $U(1)$ transformations do not commute with each other. This can be easily seen by considering infinitesimal transformations. 
In consequence, there is yet another global symmetry, 
\[
\Psi_{\beta}= \cosh\beta\: \Psi - i \:\sinh\beta\: \Psi^*,   \;\;\; \Psi_{\beta}^*=  i \:\sinh\beta\: \Psi  + \cosh\beta\: \Psi^*,  
\]
where $\beta$ is arbitrary real constant. Generators \footnote{We use the physicists convention, i. e., there is $i$ in front of real structure constants in the Lie algebra of generators} of all three global symmetries, in the order of appearance in this Section,   are isomorphic to $-i \sigma_1, \:\sigma _3,  \:-i \sigma_2$, where 
$\sigma_k$ are the Pauli matrices. They form the Lie algebra of the Lorentz group $SO(2,1)$ in the three dimensional Minkowski space-time. Let us remind that we are considering the internal symmetries, the space-time coordinates $x^{\mu}$ are not transformed.

\section{The matrices $r(x)$ of higher ranks}

We shall see below that matrix  $r(x)$ of rank $k>1$ can be written as a sum of $k$  rank one matrices of the form (13) \footnote{Formulas of this kind are well known in linear algebra as  rank decompositions, see, e.g.,  \cite{11}, \cite{12}. For convenience of the reader,  we  present  a simple derivation in the case of $k=4$}.  Therefore, we can use the results of the previous Section. We find a multiplet of $k$ independent Dirac fields. The internal symmetry group is $GL(k,\mathbb{R})$. Obviously, $k \leq 4$. For clarity, below we explicitly consider the case $k=4$. Modifications needed when $k=2$ or $3$ are obvious.

It the case at hand, formula $(A1)$ from the Appendix A reads
\begin{equation}
r = \left(\psi_1, \psi_2, \psi_3, \psi_4 \right)\: \left(\chi_1, \chi_2, \chi_3, \chi_4 \right)^T,
\end{equation}
where $\psi_i, \chi_i$, $i=1, \dots 4$,  are  four-component Majorana bispinors. 
Let us write the matrix $\left(\psi_1, \psi_2, \psi_3, \psi_4 \right)$  as the sum of matrices of rank 1  (or rank 0 if certain $\psi_i=0$),
\begin{equation}
\left(\psi_1,  \ldots \psi_4 \right) = \left(\psi_1, \bf{0}, \bf{0}, \bf{0} \right) +  \left(\bf{0}, \psi_2, \bf{0},  \bf{0} \right) +   \left(\bf{0}, \bf{0},  \psi_3, \bf{0} \right) + \left(\bf{0}, \bf{0}, \bf{0}, \psi_4 \right), 
\end{equation}
where $\bf{0}$ denotes the Majorana bispinor with vanishing all four components. 
Next, we use formulas 
\[ \left(\psi_1, \bf{0}, \bf{0}, \bf{0} \right)\: \left(\chi_1, \chi_2, \chi_3, \chi_4 \right)^T  = \psi_1\: \chi_1^T, \;\;\;\; \left( \bf{0}, \psi_2,\bf{0}, \bf{0} \right)\: \left(\chi_1, \chi_2, \chi_3, \chi_4 \right)^T  = \psi_2\: \chi_2^T,  
\]
and so on. Therefore,
\begin{equation}
 \left(\psi_1,  \psi_2, \psi_3,  \psi_4 \right)\: \left(\chi_1, \chi_2, \chi_3, \chi_4 \right)^T
= \sum_{i=1}^4 \psi_i \: \chi_i^T. 
\end{equation}
Each pair $\psi_i, \chi_i^T$  on the r.h.s. of this formula forms the Dirac field as shown in Section 3.  Thus, we find a multiplet of four independent Dirac fields. 

The matrix $r$ has 16 real matrix elements while on the r.h.s. of formula (28) there are 32 independent real matrix elements. This discrepancy is superficial, because there is $GL(4, \mathbb{R})$ gauge invariance of the form
\begin{equation}
\psi'_s(x) = \psi_i(x) G_{is}(x), \;\;\; \chi'_s(x) = \chi_k(x) \left(G^{-1}(x)\right)_{sk}, 
\end{equation}
where the $G(x) \in GL(4, \mathbb{R})$. The new fields $\psi_i', \chi_i'$ give the same matrix $r$ as $\psi_i$ and $\chi_i$. 
The gauge  transformation (31) can be written in  compact matrix form as
\begin{equation}
\left(\psi'_1,  \psi'_2, \psi'_3,  \psi'_4 \right)(x) =   \left(\psi_1,  \psi_2, \psi_3,  \psi_4 \right)(x)\: G(x), 
\end{equation}
\begin{equation} 
   \left(\chi'_1, \chi'_2, \chi'_3, \chi'_4 \right)(x) = \left(\chi_1, \chi_2, \chi_3, \chi_4 \right)(x)\:\left(G^{-1}(x)\right)^T. 
\end{equation}
The local  $GL(4, \mathbb{R})$ group contains 16 arbitrary real functions which can be used to eliminate 16 superfluous components on the r.h.s. of formula (28). Note that transformation (33) of the fields $\chi_i$  is contragredient conjugate to transformation (32) of the fields $\psi_i$.

In general, the number of Dirac fields equals $k = rank\: r$, and the gauge group is $GL(k, \mathbb{R})$. 

Similarly as in the case $k=1$ discussed in the Section 3, one can consider  conserved currents and non-Abelian gauge fields related to the $GL(k, \mathbb{R})$  gauge group. 
We shall not delve into this rather complex topic here. 

\section{Summary and remarks}

We have shown that matrix factorization of generic real $Spin(3,1)$ tensor of degree two provides the classical Majorana and Dirac fields. These fields appear as a kind of coordinates in the Lorentz invariant subsets of the full space of such tensors. The subsets are
characterized by the fixed rank $k$,  $\:1 \leq k \leq 4$,  of the matrices $r$ representing the tensors. It is a new perspective on the fermionic fields which, in particular, exposes  the local $GL(k, \mathbb{R})$ gauge transformations as their intrinsic property.  The global version of these transformations becomes a symmetry of the Lagrangian for the free Dirac fields.  In the case of $k=1$, all this, including the corresponding conserved current, has been discussed in detail in Section 3. It turns out that the internal global  symmetries form the $SO(2,1)$  group. The local $GL(1, \mathbb{R})$ transformations are preserved as  gauge invariance of the Lagrangian provided that the Abelian gauge field $B_{\mu}$ and  pertinent gauge covariant derivatives are  introduced, as shown also in the Section 3.  This gauge invariance exists in parallel with the well known global $U(1)$ symmetry, which gives rise to the $U(1)$ gauge invariance after introducing the corresponding gauge field $A_{\mu}$.

The present work can be continued in at least three directions. First, we did not discuss in detail the three cases $k=2, 3, 4$. Here  we have   multiplets of $k$ Dirac fields, and  the symmetry group $GL(k, \mathbb{R})$ is non-Abelian. Thus, the resulting gauge invariant model is rather complex. Moreover, the  $GL(k, \mathbb{R})$  group is non compact. Its dimension is equal to $k^2$, thus there will be $k^2$ four-vector real-valued gauge fields. All this may significantly complicate  construction of a satisfactory model.  On the other hand, these models could be  rather special in the sense that the  $GL(k, \mathbb{R})$ group is the truly intrinsic symmetry of the fields, as shown above. 

Another topic deserving further investigation is the role of the  $GL(k, \mathbb{R})$  symmetry in the quantum theory of the free  Dirac field(s). We expect that the  $GL(1, \mathbb{R})$ charge will not commute with the $U(1)$ charge.
It could happen that the vacuum state for the free Dirac field can not be both $GL(1, \mathbb{R})$ and $U(1)$ neutral. 

 One could also think of a quantum model parallel to QED, with the  $GL(1, \mathbb{R})$ gauge field  $B_{\mu}$ coupled with the current $j^{\mu}_s$ given by formula (20). One may guess that the Feynman perturbative amplitudes in such a model will differ from the ones of QED. Perhaps one could construct also quantum models with the non-Abelian gauge fields of the  $GL(k, \mathbb{R})$ type with  $k=2,3$, or $4$.

\section*{Appendix A. \; Rank of matrices}

Here we recall mathematical facts about rank of a matrix which we use our paper. 
Precise definitions, proofs, and more information can be found in, e.g., \cite{10}, \cite{11}. Very useful can be the article in Wikipedia \cite{12}. 

Let $M$ be a real matrix with $m$ rows and $n$ columns. The columns and rows can be regarded as $m$ dimensional or $n$ dimensional real vectors,  respectively.
 In our paper $m=n=4$. 
 The rank of $M$, denoted $rank\: M$, is defined as the maximal number of linearly independent columns of $M$. It is equal to the maximal number of linearly independent rows of $M$.  
 
  Let $A$ and $B$  be  nonsingular square matrices of the size $m$ by $m$ or $n$ by $n$, respectively. Then $rank\:(AM) = rank\:(MB) = rank\: M. $ Thanks to this property rank of the matrix $r(x)$ is Lorentz invariant.

In the case $rank\: M=1$,  $i$-th column of $M$ has the form $c_i \psi$, where $\psi$ is arbitrary non vanishing column of $M$ and $c_i$ are real numbers. Therefore,
\[ M= \left( c_1  \psi, c_2 \psi, \ldots c_n \psi \right) = \psi\: \chi^T, \;\;\;\;\;\; \chi= \left(\begin{array}{c} c_1 \\ c_2\\ \vdots \\ c_n \end{array}  \right). \] 

In general, if $rank\: M =k >0$, the matrix $M$ can be written in the form
\[ M = \left(\psi_1, \psi_2, \ldots \psi_k \right)\: \left(\chi_1, \chi_2, \ldots \chi_k \right)^T,    \eqno{(A1)}\]
where $\psi_1, \ldots \psi_k$ is the maximal set of linearly independent columns of $M$ and $\chi_1, \ldots \chi_k$ are certain $n$ dimensional vectors (represented as columns).

\section*{Appendix B. \;  The proof of conditions (12)}

The matrix of rank zero, i.e.,  $\tilde{r}(x)=0$, trivially obeys  conditions (12). 
Therefore we assume  below that $\tilde{r}(x)\neq 0$. 

Let us multiply all formulas in (12) by $\tilde{r}(x)^T$ from the left, and by   $\tilde{r}(x)$ from the right. Introducing the symmetric, real  matrix $X= \tilde{r}(x)^T\: \tilde{r}(x)$ we have the conditions
\[
X \gamma^0 X=0, \;\;\;  X \gamma_5 \gamma^0 X=0, \;\;\;  X \gamma_5 X=0, \;\;\; X \gamma^{k} \gamma_5 \gamma^0 X=0, \eqno{(B1)}
\]
where $k=1,2,3$. Note that the matrix $X$  is not vanishing because $\tilde{r} \neq 0$.  Hence, it possesses at least one real eigenvalue $\lambda\neq 0$ and the corresponding real normalized eigenvector $e$, $Xe = \lambda e$, $e^T e =1$. 
The first three conditions $(B1)$ give
\[
X \gamma^0 e=0, \;\;\;  X \gamma_5 \gamma^0 e=0, \;\;\;  X \gamma_5 e=0.    \eqno{(B2)}
\]
Let us consider the real vectors $f_1= i\gamma^0e, \;f_2= \gamma_5 \gamma^0 e, \; f_3= i\gamma_5 e$. They are orthogonal to each other (because of antisymmetry of  the involved matrices) and normalized to 1, $f_j^{\:T} f_k = \delta_{jk}.$  We see from $(B2)$
that they are eigenvectors of the matrix $X$ with the eigenvalues equal to 0. In consequence, the spectral decomposition of the matrix $X$ has the form $ X= \lambda \:e\: e^T$. Moreover, $\lambda >0$ because  the definition of the matrix $X$ implies that $\mbox{Tr} X >0$.

 Note that the matrix $X = \lambda e e^T$  satisfies
 the last condition $(B1)$ as a trivial identity because the matrices $\gamma^k \gamma_5 \gamma^0$, $k=1,2,3,$ are antisymmetric.

The matrix $X$ can be diagonalized, $ O X O^T = \mbox{diag}(\lambda, 0, 0, 0)$, where the matrix $O$ is a real and orthogonal.   
Thus, denoting $Y= \tilde{r}\: O^T$, we have the following equation for the four by four real matrix $Y$
\[ Y^T\: Y = \lambda \: \left( \begin{array}{cccc} 1 &0&0&0\\ 
0&0&0&0 \\  0&0&0&0\\ 0&0&0&0 \end{array} \right).  \eqno{(B3)}
\] 
Its solution has the form
\[ Y= \left( \begin{array}{cccc} h_1 &0&0&0\\ 
h_2&0&0&0 \\  h_3&0&0&0\\ h_4&0&0&0 \end{array} \right),   \eqno{(B4)}
\] 
where the first column
\[ h=  \left( \begin{array}{c} h_1\\ 
h_2 \\ h_3\\ h_4 \end{array} \right)  \eqno{(B5)}
\] 
is real, normalized to $\lambda$, i.e. $h^T\: h = \lambda$, and otherwise arbitrary. 
Thus, we have proved that the matrix $Y$ has the rank equal to 1. 

The matrix $\tilde{r}$ is obtained from the formula $\tilde{r} = Y O$, where $O$ is orthogonal, hence nonsingular.  Therefore, all  columns of $\tilde{r}$ are given by $h$ multiplied by  real numbers of which at least one is different from 0.  This means that $rank \:\tilde{r} =1$. 

The proof in the opposite direction is simple. Each matrix $\tilde{r}$ constructed from the matrix $r$ of the form (13) obeys constraints (12) trivially, because  the matrices $\gamma^0, \: \gamma_5, \: 
\gamma^{\mu} \gamma_5 \gamma^0$ are antisymmetric. 

\end{document}